\documentclass{article}

\usepackage{arxiv}

\usepackage[utf8]{inputenc} 
\usepackage[T1]{fontenc}    
\usepackage{hyperref}       
\usepackage{url}            
\usepackage{booktabs}       
\usepackage{amsfonts}       
\usepackage{nicefrac}       
\usepackage{microtype}      
\usepackage{lipsum}
\usepackage{graphicx}
\usepackage{amsmath}
\graphicspath{ {./images/} }

\title{A critical phase transition in bee movement dynamics can be modeled using a 2D cellular automata}

\author{
 Ivan Shpurov \\
  Embodied Cognitive Science Unit (ECSU)\\
  Okinawa institute of Science and Technology\\
  Japan, Onna-som 15213 \\
  \texttt{ivanshpurov@gmail.com} \\
   \And
 Tom Froese \\
  Embodied Cognitive Science Unit (ECSU)\\
  Okinawa institute of Science and Technology\\
  Japan, Onna-som 15213 \\
  \texttt{TOM.FROESE.oist.jp} \\
}

\begin{document}
\maketitle
\begin{abstract}
The collective behavior of numerous animal species, including insects, exhibits scale-free behavior indicative of the critical (second-order) phase transition. Previous research uncovered such phenomena in the behavior of honeybees, most notably the long-range correlations in space and time. Furthermore, it was demonstrated that the bee activity in the hive manifests the hallmarks of the jamming process.  We follow up by presenting a discrete model of the system that faithfully replicates some of the key features found in the data - such as the divergence of correlation length and scale-free distribution of jammed clusters. The dependence of the correlation length on the control parameter - density is demonstrated for both the real data and the model. We conclude with a brief discussion on the contribution of the insights provided by the model to our understanding of the insects' collective behavior.
\end{abstract}


\section{Introduction}
Recent decades saw a rapid diffusion of methods and concepts from the domains of physics and mathematics into biology, social sciences \cite{jusup2022social}, and other areas previously dominated by qualitative and descriptive methods. The study of collective behavior, which thought the history of human thought has been both inspirational and enigmatic, has been especially enriched by the insights from physics \cite{cavagna2008new}. 

A prominent phenomenon that collective groupings exhibit is the emergence of large-scale behavior patterns from local group interaction without a central coordinator. Periodic oscillations in the movement activity of ant colonies which supersede the spontaneous movement of individual ants \cite{cole1991short}, the emergence of polarized order in the aggregations of marching locusts \cite{buhl2006disorder}, and metastability of the fish shoals, that enables the rapid change of configuration in response to external perturbation \cite{couzin2003self} are just a few examples of decentralized collective behavior. 

Seminal theoretical works \cite{chialvo2010emergent, bak2013nature} established the crucial role critical dynamics play in the function of complex adaptive systems. In the vicinity of the critical point, correlation length and susceptibility are at their peaks, maximizing the system's responsiveness to environmental stimuli and propagation of information within the system. Concerning the collective behavior of swarms, hallmarks of criticality, such as scale-free event size distribution, long-range correlation, and scaling of susceptibility and the correlation length with the size of the system have been found in fish shoals \cite{puy2024signatures}, flocks of starlings \cite{cavagna2010scale}, and swarms of midges \cite{attanasi2014collective}. 

\begin{figure}[h]
    \centering
    \includegraphics[width=0.9\linewidth]{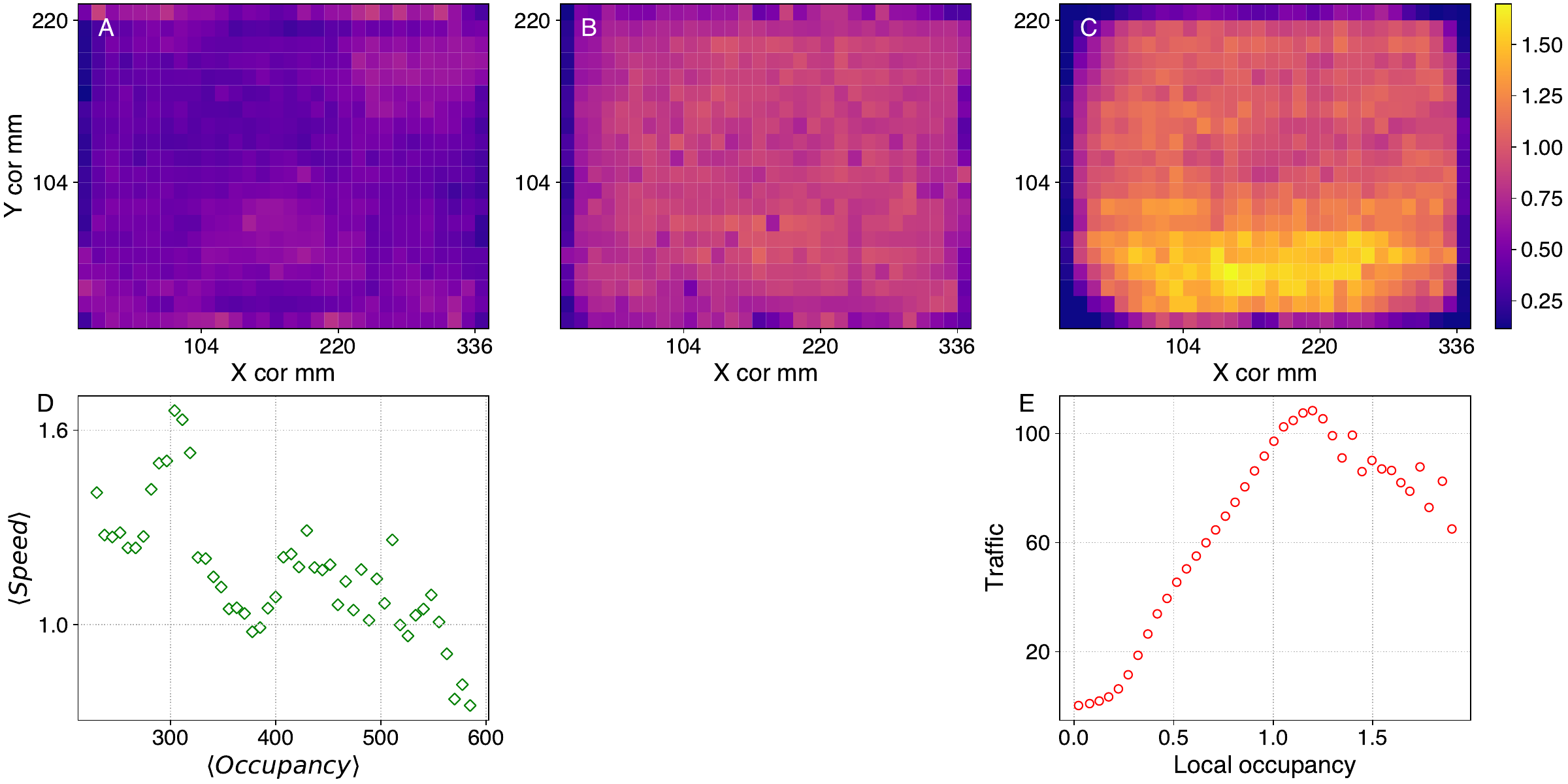}
    \caption{Data overview. Top panes A, B, and C show occupancy heatmaps of the beehive computed for 3  different average occupancy (bee count) levels: low $\approx 281$ (Pane A), medium $\approx 469$ (Pane B), and high $\approx 585$ (Pane C). Each heatmap is averaged over a two-hour interval. Pane D presents the relationship between mean hive occupancy and the mean speed of bees. Note the evident downward trend. Pane E depicts the non-linear relationship between traffic and local occupancy, evidence of the jamming process (Adapted from previous work \cite{shpurov2024beehive}). }
    \label{fig:Bees_introduction}
\end{figure}

In our previous work \cite{shpurov2024beehive}, we have established that the bee-hive exhibits scale-free correlations both in space and in time. Furthermore, our research showed that the beehive undergoes a jamming transition. The latter has been discovered in the study of traffic dynamics. Foundational work combining modeling efforts and real data work \cite{nagel1992cellular} laid out the key characteristics of the process: agents' speed is inversely related to their density, while traffic, a variable that quantifies the number of agents passing through a unit of space per unit of time, has a non-linear dependence on density. For low densities, traffic increases monotonously with density, however once the critical density is reached, further increases in density lead to a reduction in traffic. The graphical representation of this non-linear dependency has a recognizable parabolic or tent-like shape and is known as the fundamental traffic diagram. Looking ahead, one can easily identify this relationship presented on the pane E of the Figure \ref{fig:Bees_introduction}, which gives a summary of the beehive data.

To expand on our earlier results we develop a 2-dimensional jamming model: a stochastic cellular automaton, capable of replicating the main properties of the data. Using simple models to understand complex phenomena is an approach that has been exceedingly successful in physics and has proven its efficiency in the study of collective animal behavior \cite{couzin2002collective, ginelli2015intermittent}. A key prerequisite of a successful model is the choice of adequate metrics to compare the model with the real data. As our work focuses on the critical behavior in the vicinity of the phase transition, we primarily use thermodynamic metrics, such as correlation length and susceptibility. 

Jamming transition, out of the context of insect behavior, was actively investigated in simulations. Initial works focused on 1D models with agents moving along parallel lanes and were able to replicate a rich variety of emergent phenomena found in real-world systems, including "phantom" traffic jams that form in the absence of a bottleneck, and traffic hysteresis that manifests itself in different behaviors during acceleration and deceleration \cite{treiterer1974hysteresis,zhang1999mathematical}. Beyond 1D, jamming processes were studied using grid lattices which use a structure of pathways and intersections mimicking a cityscape to channel the flow of agents \cite{horiguchi1998numerical}.  Other simulation works used biased agents whose mobility was unrestricted and showed a power-law distribution of 2D jams emerging at certain parameter combinations \cite{tadaki1997distribution}. The latter interlinks with the research concerning the Asymmetric simple exclusion process(ASEP), a standard model for transport in non-equilibrium statistical physics \cite{ding2018analytical}. However, to our knowledge, the diversity of existing models has not been applied to collective animal behavior, highlighting the need for our modeling efforts.

\section{Results}
\label{sec:Redults}
This section is structured as follows: We begin with a brief overview of the data, then present a description of the model before proceeding to present the results of various simulations that show how key characteristics of the data are replicated in the model. 
\subsection{The data}

For our study, we use a dataset of bee trajectories recorded by \cite{gernat2018automated} and graciously shared with the current paper's authors. This data has already been used in our previous work \cite{shpurov2024beehive}. The recording lasted multiple days and nights: 1200 one-day-old tagged worker bees were used, along with a queen (unmarked and absent from the data). Each insect was marked with an individual barcode before being placed in a hive of dimensions 348$\times$232 mm with a transparent cover enabling constant recording. The recorded footage was then processed, and individual barcode detections were stored. Subsequently, trajectories in 1-sec resolution were reconstructed from the barcode detections. An important feature of the experimental protocol is that for the first 2 days and 2 nights of the experiment, insects were kept locked in the hive, then at sundown of the second day, the entrance to the hive was opened and insects were permitted to go out and forage. Finally, during the last 3 days and 2 nights of the experiment, foragers were not permitted to return to the hive. Besides the changes in the number of bees due to foraging, additional fluctuations in the occupancy of the hive occur as bees could go inside the honeycomb cells to take care of the brood, eat, and sleep - the latter behavior is especially common during the night time. Therefore, the average occupancy level of the hive varied significantly during the experiment. To illustrate this the top panes of Figure \ref{fig:Bees_introduction} present heatmaps of the bee distribution in the hive for 3 different average occupancy levels. 

Heatmaps are created using coarse-grained occupancy data, which is used throughout the entire work. Coarse-graining is produced by partitioning the entire hive area into 30$\times$20 square cells with a side of 11.6 mm. Such an approach permits the computation of correlations between occupancy fluctuation time series of different grid cells.

\subsection{The Model}
To construct a minimalist model of the jamming-unjamming process in the beehive we used a 2-dimensional stochastic cellular automaton with a periodic boundary condition. At $t=0$, the grid is populated with randomly placed agents, each agent occupying a single grid site. The ratio of agents to empty grid cells is the control parameter of the system and is hereafter referred to as the model density. The behavior of the individual agents is dictated by 4 variables: bias parameter $B_{A}$, maximal movement distance $D_{max}$, sight range, and waiting time $\tau_{W}$. The bias parameter specifies a preferred direction of movement (out of 4 possible) and is bound between 0 and 1.
The agents move according to the following rules: 
\begin{enumerate}
    \item Check two directions of movement for the presence of other agents within the sight range. The direction specified by the bias parameter is picked with a probability $B_{A}$, and all remaining directions have $(1-B_{A})/3$ probability.
    \item If no agents are detected, move $D_{max}$ steps in any of the two directions.
    \item If other agents are present within the sight range, the agent moves in a direction that permits maximal movement. The agent will terminate its movement once it occupies a cell adjacent to the closest neighbor. If the distance to the nearest neighbor is the same for both evaluated directions choice is made randomly.
    \item If both directions that were chosen at step 1 are blocked (adjacent cells are occupied), the agent remains stationary, and the waiting time counter starts.
    \item If the waiting time counter exceeds $\tau_{W}$ new bias parameter for the agent is randomly chosen and will be used in the next cycle. 
\end{enumerate}
At each timestep, all agents in the model compute their movements. Then, movements that result in a clash (two agents occupying the same grid cell) are canceled, and those agents remain stationary. The waiting time counter is incremented by 1 for every step the agent spends in a stationary state and is reset to 0 once the agent moves.  

\begin{figure}[h]
    \centering
    \includegraphics[width=0.9\linewidth]{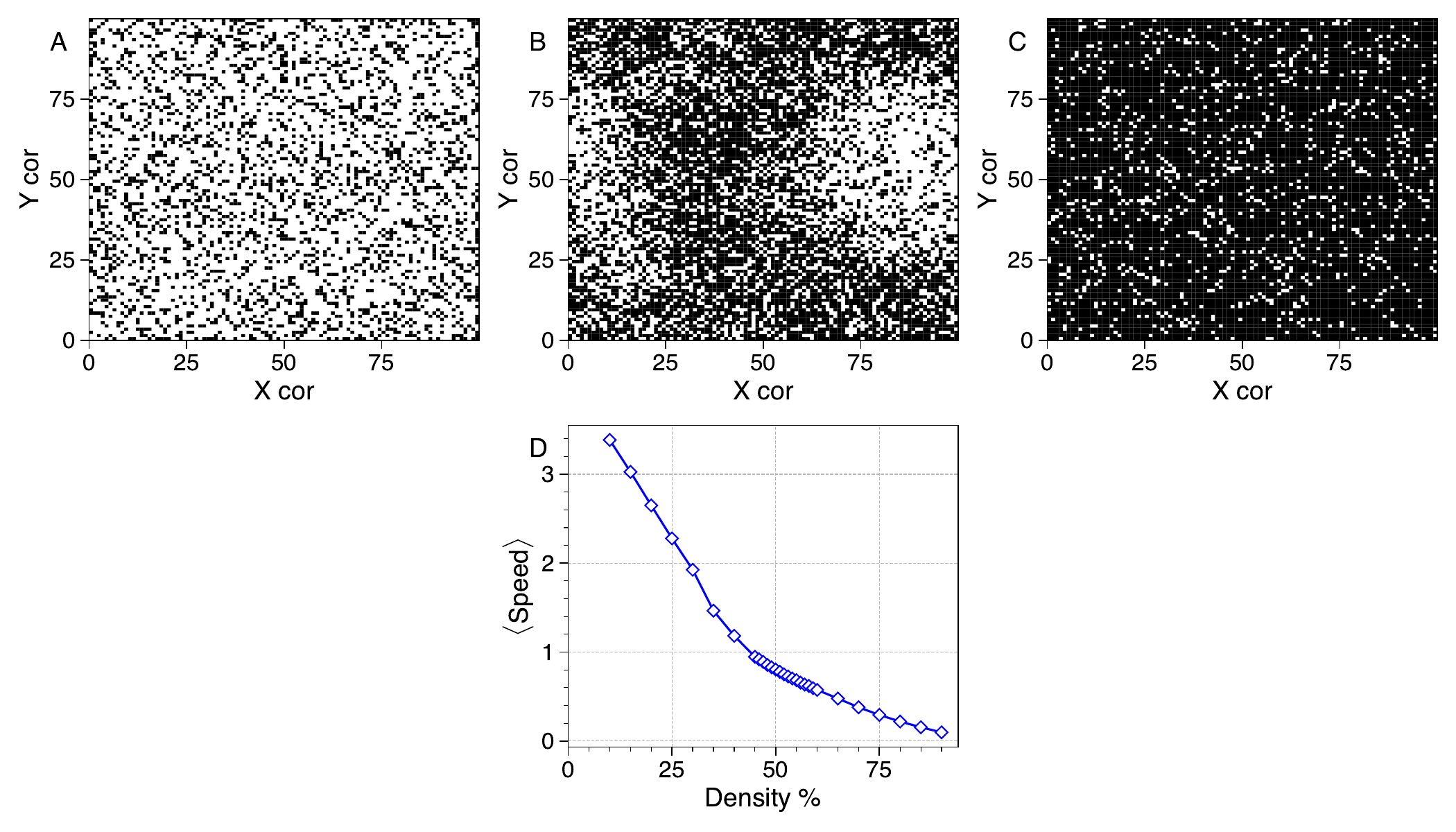}
    \caption{Model overview. Top panes A, B, and C show snapshots for different 100$\times$100 models at $t=20000$. Three densities are presented: low-25\% (Pane A), intermediate-56\% (Pane B), and high-90\% (Pane C).  Pane D shows the relationship between the average speed of agents and the model density. A clear downward trend is evident.}
    \label{fig:Model Introduction}
\end{figure}

This model draws on the previous work, most notably the classical models of 1-dimensional jamming and ant movement \cite{rauch1995pattern, nagel1992cellular} and ASEP models. From a cursory review of the rules governing agents' behavior, it is self-evident that agents' mobility is dependent on the overall density of the model. This intuition is confirmed by the inverse relationship between the average agents' speed and model density (Figure \ref{fig:Model Introduction}, Pane D). Movements of individual agents' are interdependent and give rise to interesting large-scale dynamics. 

In the current work, models of different densities and sizes are used. In all cases bias $B_{A}$parameter is set to 0.4, and the sight range and $D_{max}$ are equal to 5. The waiting time $\tau_{W}$ is set to 10. Simulation time $t=40000$.

Our model expands previous discovery of the jamming transition in the beehive. Two hallmarks of jamming  - the inverse relationship between the mean speed of agents and their density, and the non-linear relationship between the density and traffic are present in both the model and the data. Relationships between average speed and density are presented in Figure \ref{fig:Bees_introduction} (pane E) and Figure \ref{fig:Model Introduction} (Pane D) for the data and the model respectively. It should be noted that although the clear downward trend is present in both cases it is considerably noisier for the bee data. Apart from the usual ramifications, this is explained by the circadian rhythms inherent to the functioning of the beehive - during the night, both the occupancy and the average speed are lower than during the day. This consideration explains seemingly premature drops in average speed when the $\langle Occupancy \rangle$ is in the range [300, 400], or below 200 in the leftmost part of the graph (Figure \ref{fig:Bees_introduction}, Pane E).

Phase transitions are characterized by a divergence of correlation length $\xi$ when the control parameter approaches the critical point \cite{sethna2021statistical}. When the model is considered the dependence of correlation length on density can be studied straightforwardly, by instantiating multiple models with different densities. When the beehive data is considered it is possible to leverage variability in the average occupancy level to reconstruct a dataset by grouping short segments of occupancy time-series with similar average occupancy values. The reconstructed data can be then used to compute correlations for different average occupancy levels. It should be noted that we refer to the control parameter of the model as density and express it as a \% value, while the control parameter of the beehive is named \textit{Occupancy} and is expressed as the average number of bees per time interval. Such terminology is used to avoid confusion - both are control parameters; however, the model density is more precise, as it reflects a portion of the area covered by the agents. The dataset represents bees as points-masses, therefore the exact area they cover is unknown. 

\begin{figure}[h]
    \centering
    \includegraphics[width=0.9\linewidth]{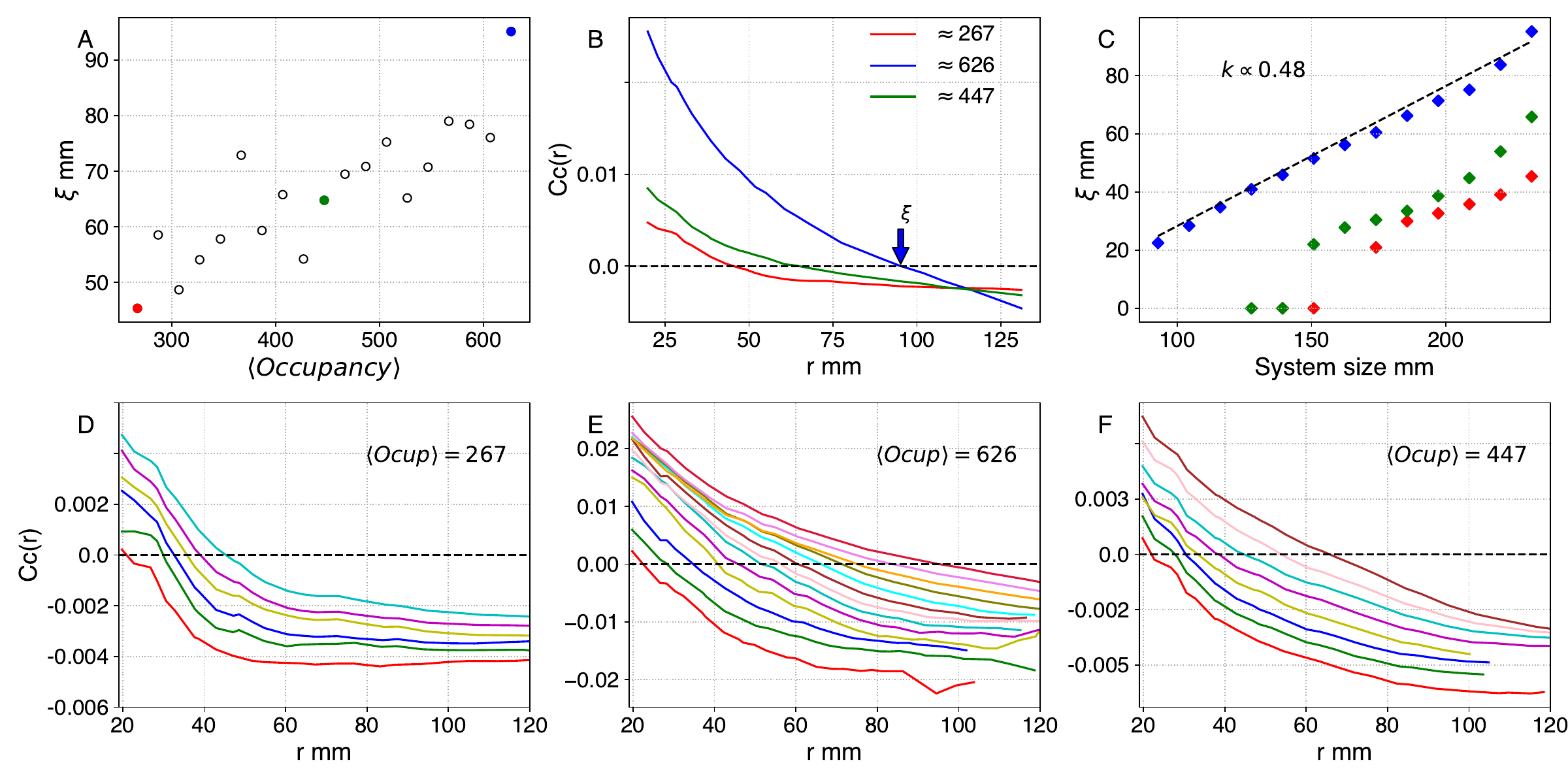}
    \caption{Connected correlation function and its scaling (Data). Pane A presents the relationship between the average occupancy of the beehive and the correlation length $\xi$. The computation uses 30$\times$20 coarse-graining. Pane B presents connected correlation functions for 3 different occupancy levels. Pane C shows the scaling of the correlation length for subsystems of different sizes computed via box scaling. Blue diamonds correspond to the highest density; note the linear trend indicative of the critical state of the system. Panes D, E, and F show connected correlation functions for 3 subsystems of increasing size. Respective average occupancy levels are 267 (Pane D), 626 (Pane E), and 447 (Pane F).}
    \label{fig:Bee_correllations}
\end{figure}

Correlation length $\xi$ indicates the characteristic distance over which elements of the system are "aware" of each other’s states. Of particular importance is that the correlation length diverges in the vicinity of the critical point.  When dealing with dynamics of biological systems it is commonly computed as the first zero crossing of the connected correlation function Cc(r) \cite{attanasi2014collective,cavagna2010scale,fraiman2012kind}. The latter is defined as the correlation function of the fluctuations $\delta x$ around the mean of the signal $X$ \eqref{eq:mean_fluctuaitions}. 
\[
\delta x_{i}=x_{i}-\frac{1}{N}\sum_{1}^{N}x_{i}
\label{eq:mean_fluctuaitions}
\tag{1}
\]
\[
Cc(r)=\frac{\sum (\delta x_i(r)-\langle \delta x(r) \rangle)(\delta y_i(r)-\langle \delta y(r) \rangle)}
{\sqrt{\sum (\delta x_i(r)-\langle \delta x(r) \rangle)^2}\sqrt{\sum(\delta y_i(r)-\langle \delta y(r) \rangle)^2}}
\tag{2}
\label{eq:Space_correlations}
\]

We use the occupancy time series of the coarse-grained beehive data and the model. After the mean is subtracted from the beehive data, distance-binned correlation functions are computed for different occupancy levels using \eqref{eq:Space_correlations}. Note, that step \eqref{eq:mean_fluctuaitions} is unnecessary when the model timeseries are considered as density remains stationary during the simulation. Correlations are computed in the same manner for the model data.

Panes B in Figures 3 and 4 show the shapes of the connected correlation functions for the data and the model at 3 different densities (occupancies). Zero crossings are marked with arrows. The evident dependence of the correlation length on the control parameter is further illustrated by Panes A in the same figures, which shows how $\xi$ changes for models of different densities as well as for different occupancy conditions of the beehive. 

In the model data correlation length peaks when the density is at 58 \% and decays afterward. Therefore we observe a subcritical behavior at lower densities and supercritical behavior at higher densities. When we consider the dependence of $\xi$ on average occupancy for the beehive, it is observed that the correlation length peaks at maximal density - we don't have data for beehive behavior at supercritical densities. This observation is biologically relevant and will be expounded on later in the paper.    

Additionally,  we are interested in seeing how the correlation length scales with the model size. In the case of the model, this is accomplished by considering models of different sizes. For the beehive data, we use a method known as box-scaling developed in \cite{martin2021box}. The system of size L is subdivided into subsystems -  boxes of size W (W<L). CCF is then computed for each subsystem using its mean signal to compute fluctuations \eqref{eq:mean_fluctuaitions}. Statistical averages are taken for subsystems of the same size. For the beehive, we consider isotopic partitions of 20$\times$30 coarse-graining, starting with a 92$\times$92 mm square and concluding with the largest isotropic partition of the system with dimensions 232$\times$232 mm.

\begin{figure}[h]
    \centering
    \includegraphics[width=0.9\linewidth]{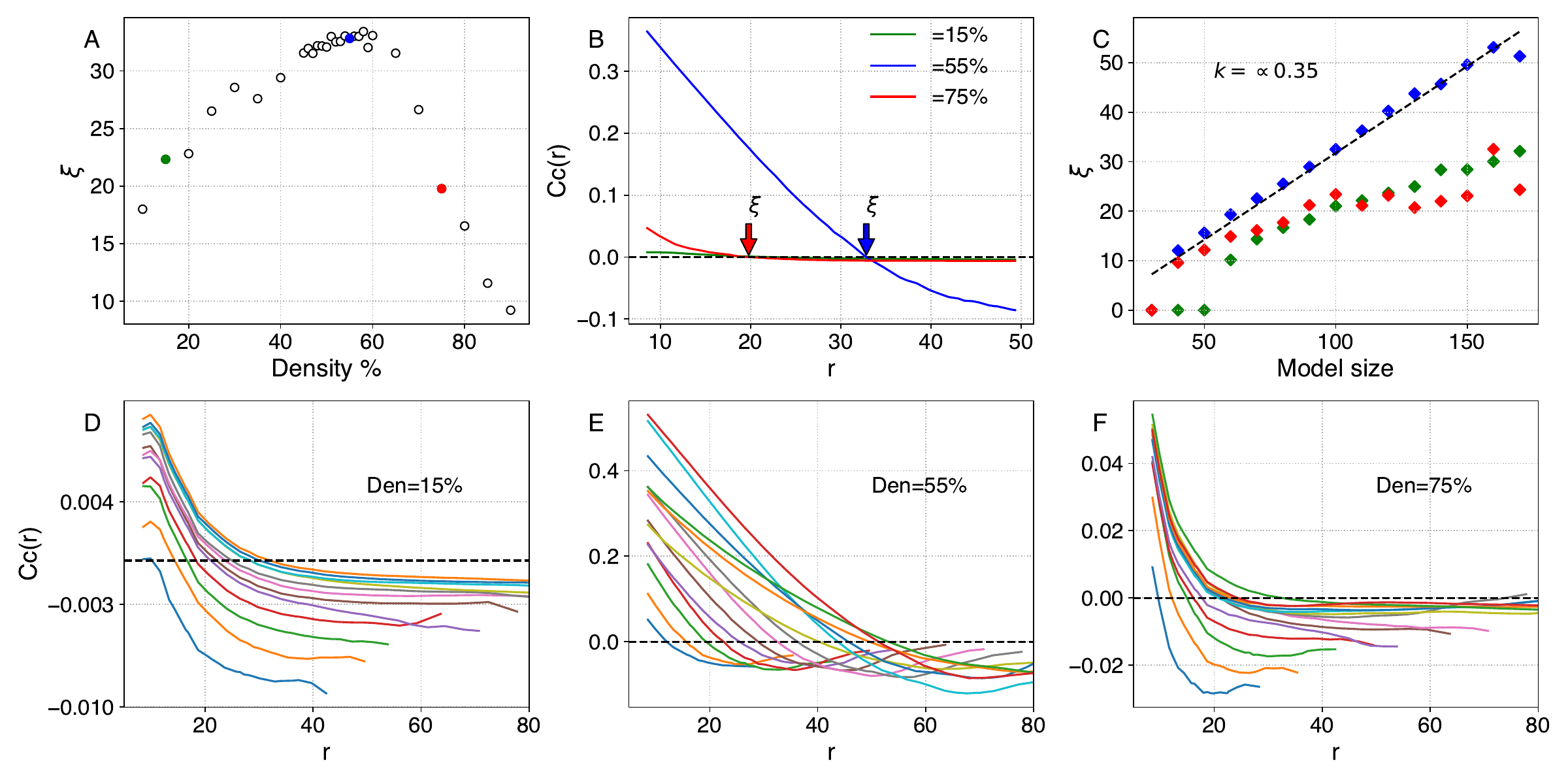}
    \caption{Connected correlation function and its scaling (Model). Pane A shows the dependence of correlation length $\xi$ on density, computed using 100$\times$100 models with different agent densities. Pane B shows 3 connected correlation functions for models with low - 15\% (green), intermediate 55\%, and high 75\%  densities. Pane C shows how the correlation length scales with model size. Markers of the same color correspond to models with the same density. Note the linear trend scaling observed for intermediate (critical) density. Panes D, E, and F show connected correlation functions for models of increasing sizes. Respective densities are 15\% (Pane D), 55\% (Pane E), and 75\% (Pane F).}
    \label{fig:Conneced_corelation model}
    
\end{figure}Pane C in Figure \ref{fig:Conneced_corelation model} shows how the correlation length scales with the model size, while pane C in Figure \ref{fig:Bee_correllations} presents the scaling of the correlation length for subsystems of different size constructed from the bee hive occupancy data. In both cases near the critical point, the scaling is linear (blue markers).  This is not the case for subcritical and supercritical densities in the model, as well as for low and intermediate occupancy levels in the beehive.

\begin{figure}
    \centering
    \includegraphics[width=0.9\linewidth]{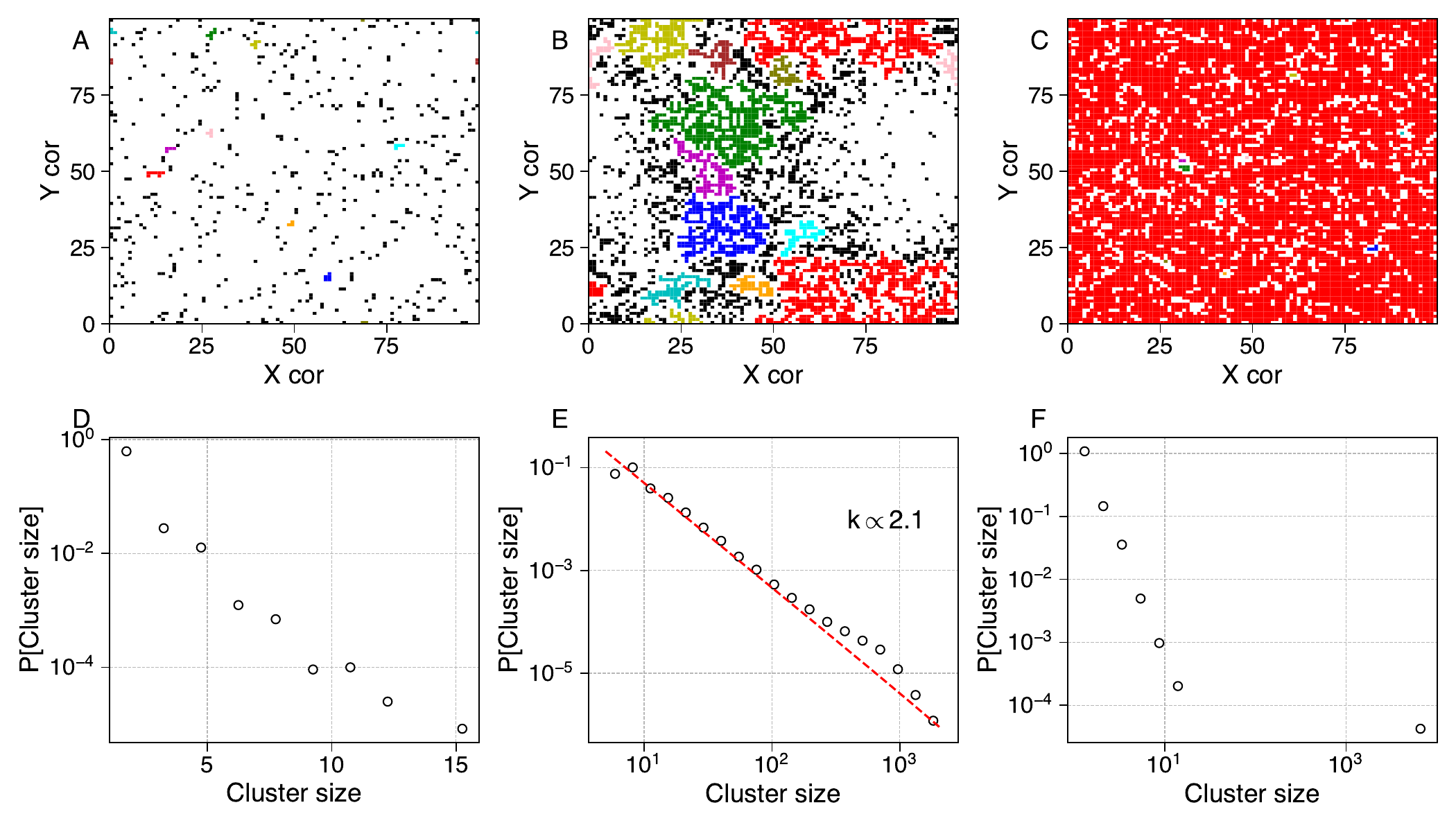}
    \caption{Cluster size distribution (Model). Panes A, B, and C present model cluster configurations formed by jammed agents at $t=20000$. Note that only immobile agents are shown. The 13 largest clusters are colored, the biggest cluster is indicated with red coloring, while all remaining clusters are black, including clusters of size one. A toroidal grid is used. Note the large variability of cluster sizes depicted on pane B (critical density). Panes D, E, and F show the probability (normalized log-binned histogram) distribution of cluster sizes for three different densities: 15\% (Pane D), 55\% (Pane E), and 90\% (Pane F). The dashed red line (Pane E) shows the maximum-likelihood fit to the power-law distribution.}
    \label{fig:Model_clusters}
\end{figure}

Additionally, we study the cluster size distribution. It is known that cluster size distribution exhibits power-law scaling at criticality. Figure \ref{fig:Model_clusters} shows the cluster size distribution for different model densities. A cluster is defined as an aggregation of immobile agents that are in contact along at least one edge. The top (Panes A, B, and C) row presents illustrative snapshots of cluster configurations while the bottom row shows statistics for cluster size distribution obtained by by counting all clusters on simulation frames between t=20000 and t=30000. Power-law is fitted to data using the max-likelihood estimator, and the quality of fit is evaluated using the log-likelihood test \cite{alstott2014powerlaw}. 

Defining clusters formed by "jammed" bees is less straightforward: the insects are capable of meandering behavior even when spatially confined. We implement a technique adapted from \cite{mashanova2010evidence} and plot displacement per second against the turning angle between successive displacements (Figure \ref{fig:Bee_clusters} Pane B). Two distinct modes of movement can be distinguished  -  meandering, with uniform turning angle distributions and lower displacement, and directed movement, characterized by higher displacement and small turning angles. Judging that meandering behavior is primarily imposed by spatial confinement by other insects, we choose to use a threshold of 4 mm/sec displacement to separate insects with relative freedom of movement from those that are part of a jammed cluster. Insects that are within distance $r$ of each other are grouped into clusters, and $r$ is set to be 10 mm to match the approximate length of a bee. Illustrative clusters configurations for low and high occupancy are shown on Panes A, and D of Figure \ref{fig:Bee_clusters}. Pane C of this figure presents two probability density distributions computed by counting all cluster sizes when the hive is at a given occupancy level. It should be noted that power-law exponents $k$ are $\approx 2$ for both the model and the beehive data near the critical value of the control parameter. 

\begin{figure}
    \centering
    \includegraphics[width=0.9\linewidth]{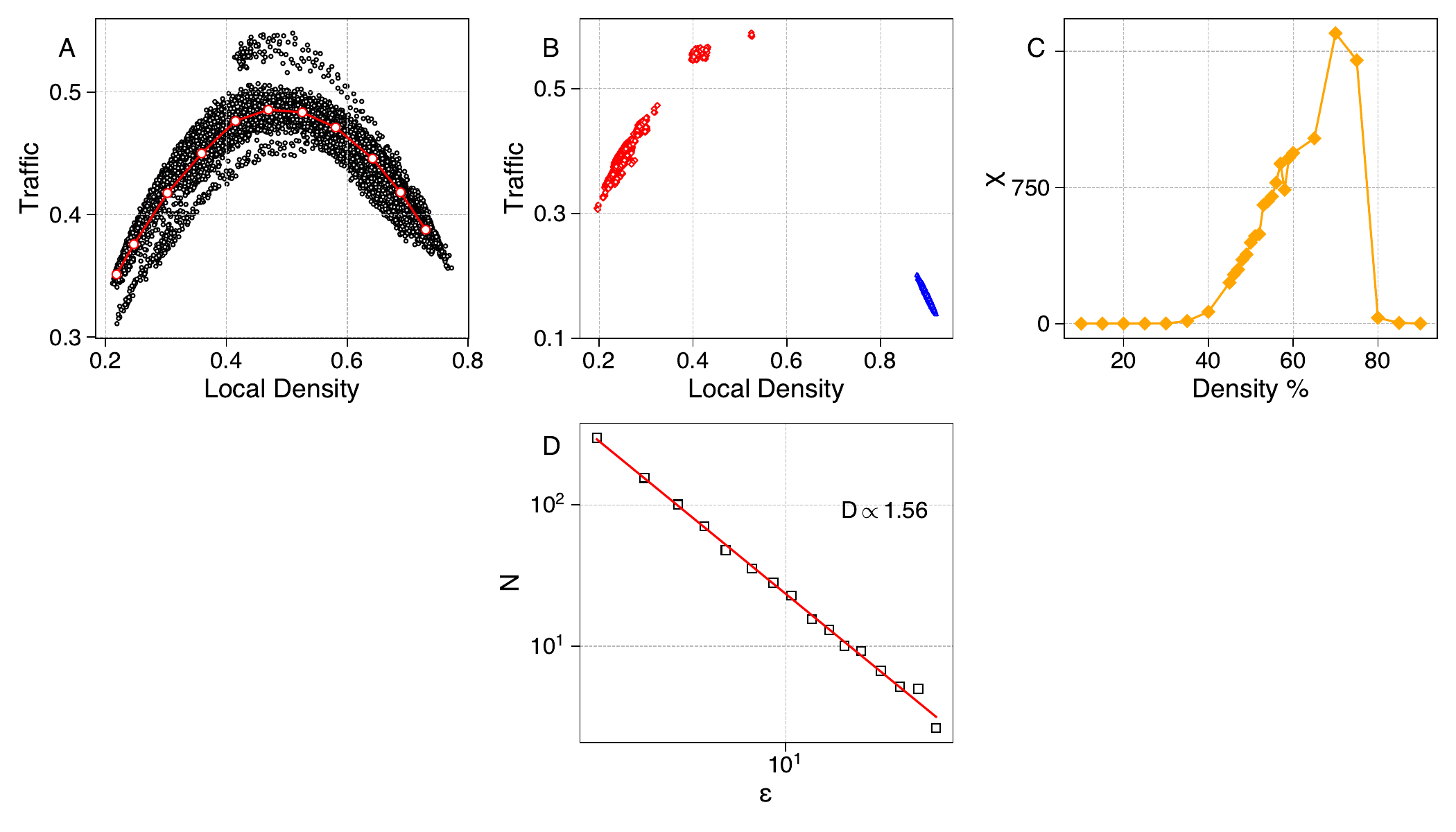}
    \caption{Pane A shows a non-linear relationship between density and traffic for models with density 55\% (critical), while pane B presents such relationships for models with density 15\%(red markers) and 90\% (blue markers). Pane C demonstrates a relation between density and susceptibility $\chi$. Pane D exhibits the box-counting dimension of the largest cluster in the critical density model.}
    \label{fig:Model_traffic}
\end{figure}

\begin{figure}
    \centering
    \includegraphics[width=0.9\linewidth]{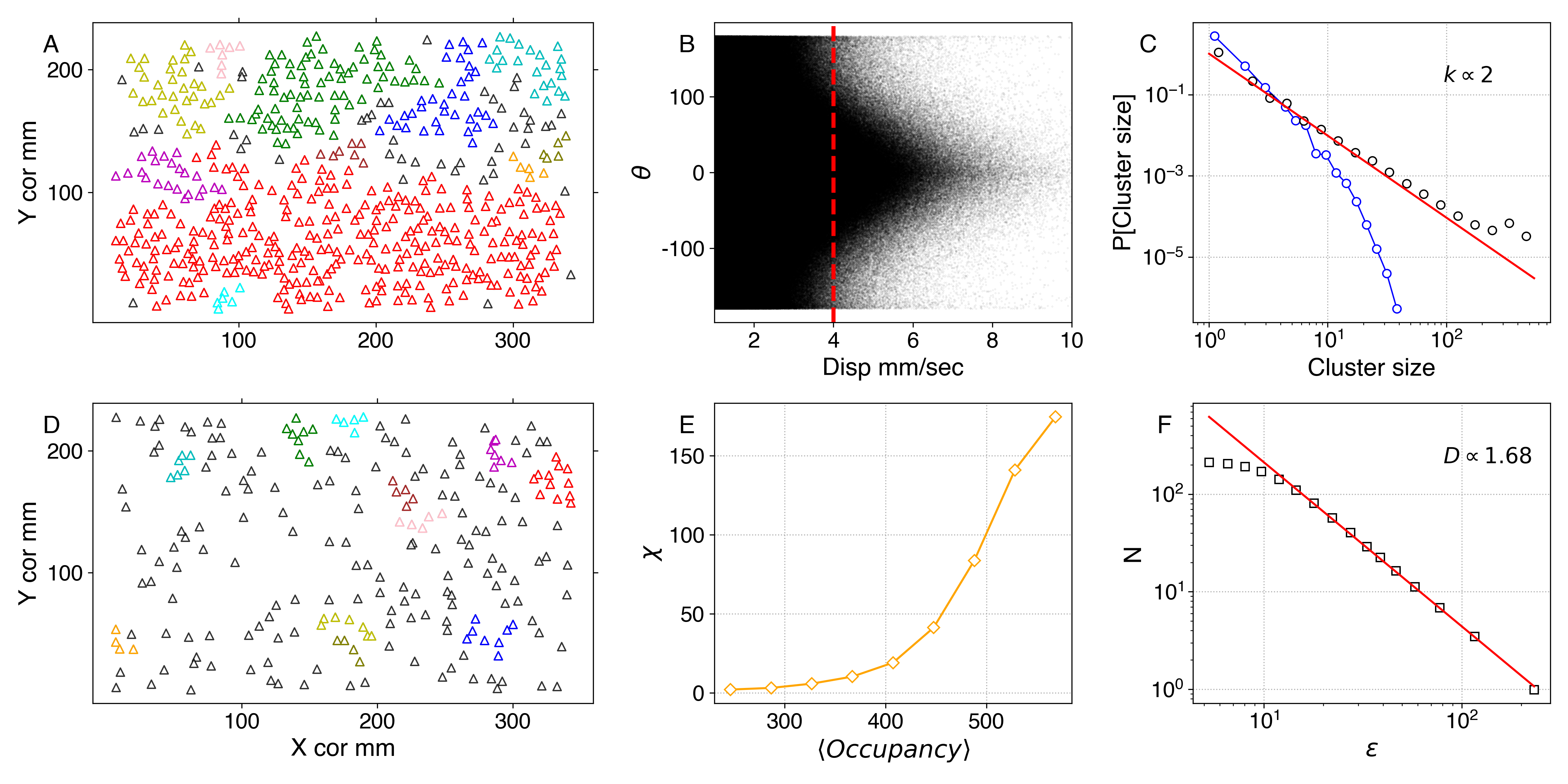}
    \caption{Clustering and fractal dimensions of the beehive. Panes A and D present snapshots of jammed agents in the beehive. The 13 largest clusters are colored, and the largest cluster is indicated with red coloring. All remaining clusters are shown with black triangles. Pane B shows the relationship between displacement per second and turning angle between successive displacements. Pane C exhibits cluster size distribution for high (black circles) and low (blue circles) hive occupancy. Pane E shows susceptibility $\chi$ as a function of occupancy, while pane F presents the computation of the box-counting dimension for the largest cluster at maximal (critical) density.}
    \label{fig:Bee_clusters}
\end{figure}

Additionally, cluster size distribution can be used to compute susceptibility $\chi$ by considering the variance of cluster sizes $s$ while excluding the largest cluster \eqref{eq:susseptability}. Results of this computation are presented in Figure \ref{fig:Model_traffic} Pane C for models of different densities and in Figure \ref{fig:Bee_clusters} Pane E for the model. The susceptibility goes down after a peak in the case of the model, but has a maximum at the highest occupancy level for the beehive.
\[
\chi=\frac{\langle s^2 \rangle - \langle s \rangle^2}{\langle s \rangle}
\label{eq:susseptability}
\tag{3}
\]

A cursory examination of the top panes of Figure \ref{fig:Model_clusters} yields that jammed clusters have complex shapes with rugged boundaries and internal lacunae. Their complexity could be quantified by considering their fractal dimension \cite{mandelbrot1983fractal}, computed via box-counting. Figure \ref{fig:Model_traffic} Pane D shows a box-counting dimension of the largest cluster. Similar computations have been made using a set of points that comprises the largest cluster of jammed bees and are shown in Figure \ref{fig:Bee_clusters} Pane F.


\section{Discussion}

Our results show that a minimalistic cellular automata model is capable of replicating some key features of the large-scale dynamics of the beehive - most notably the divergence of correlation $\xi$ length at a critical value of the control parameter, dependence of the $\xi$ on density (occupancy) and scale-free distribution of cluster sizes in the vicinity of the critical point with the exponent $k \approx 2$. The value of the exponent matches that of the percolation transition. The fractal dimensions $D_{f}$ of the largest clusters for both the model and the hive data are lower than $D_{f}\approx 1.9$  expected of the percolation on the 2D lattice \cite{shante1971introduction}. Both the model and the beehive also exhibit the hallmarks of the jamming process.   

Criticality recently started to emerge as a key universal property of complex adaptive systems capable of emergent behavior \cite{chialvo2010emergent}. Scale-free phenomena associated with criticality have been discovered in systems that are paradigmatic examples of complex collective behavior \cite{attanasi2014collective, puy2024signatures, cavagna2010scale}. However, although the work is ongoing, currently we lack elucidation of generative mechanisms underlying critical behavior that are tailored to specific cases. Our work seeks to bridge this gap through the use of a very simple model that replicates key properties of the data.

The collective behavior of honeybees is renowned for its complexity \cite{seeley2011honeybee}, and by no means can our model veridically capture all of it. A non-exhaustive list of the rich behavior repertoire purposefully omitted from the model includes circadian rhythms, trophallactic interactions \cite{gharooni2024computational, gernat2018automated}, the special role of foragers \cite{doi2023spontaneous}, and the functions of the queen. However, we don't see that as a shortcoming of the model -  on the contrary, it is remarkable that scale-free behavior can be replicated using a very simple set of behavioral rules.  In our view, criticality doesn't require complex structures to emerge but is itself a necessary foundation for the emergence of intricate behaviors. 

Several elements of the results section deserve a more detailed discussion. As noted previously the beehive data lacks a supercritical state which is present in the model. Bees have a means of controlling the occupancy level of the hive, even when it is closed, by going inside the honeycomb grid. Indeed, even at its maximum, the average occupancy is less than the total number of insects. We conjecture two factors which are applicable here: thermoregulation is a known problem for large bee aggregates \cite{peters2019collective, peters2022thermoregulatory}, therefore, supercritical occupancy levels might be avoided by insects as they would lead to an uncomfortable and dangerous increase in temperature. The alternative hypothesis that doesn't necessarily invalidate the previous conjecture is that bees are capable of controlling the distance to the critical point depending on the situation. Such ability has been hypothesized for collective entities \cite{romanczuk2023phase}, and modeling work \cite{chialvo2020controlling} showed that such regulation could be achieved using local information.

A somewhat puzzling observation is that, for the model,  a peak in susceptibility, computed using cluster size variance is shifted to the right as compared to the peak correlation length determined as the zero crossing of the connected correlation function. The obvious explanation is that the computation of connected correlation incorporates all of the data, while susceptibility is computed using only the agents that constitute the jammed clusters. The alternative hypothesis we entertain is that the system may exhibit two interacting phase transitions acting at different temporal scales - jamming transition mediated by interactions between individual agents and percolation transition, driven by the formation of clusters of jammed agents. Whether the same is true for the beehive remains to be seen: in the absence of supercritical behavior, it is unclear if the observed peaks in $\chi$ and $\xi$ are the maximum attainable for the systems, or if they would increase further with the increase in occupancy.

Finally, it should be noted that we don't pose our model and jamming process as the universal explanation for all scale-free phenomena observed in collective systems. Bees are characterized by small nearest neighbor distances that might be a necessary prerequisite for the jamming process. Staling flocks and swarming midges also exhibit scale-free correlations and the divergence of correlation length, however their interaction distances are considerably larger. It remains to be discovered if they undergo some form of jamming or if, for such systems critical state is mediated by an entirely different set of mechanisms. Other systems, such as clusters of migrating bacteria or dense human aggregates, which have short interaction ranges and manifest scale-free behavior, should be examined for signs of the jamming process.

\section{Conclusion}
This work presents a minimal model of the critical state in the beehive and shows how critical phenomena in a living collective system can arise from the most basic interactions.  
The presented findings pose new questions, be addressed in further research. Crucially, other collective systems exhibit scale-free phenomena and are characterized by short interaction ranges between neighbors should be examined for hallmarks of the jamming transition. Furthermore, the conspicuous absence of a supercritical state in the beehive requires further research to confirm, using larger datasets and more nuanced recording methods, which are now becoming available\cite{ulrich2024autonomous} to establish if the aforementioned absence is a limitation of the data at hand or a fundamental quality of bee collective behavior, which emerges from either from thermoregulation requirements or adaptive tuning of the system to the critical point. 

\section{Acknowledgments}

The authors thank Professor Dante R. Chialvo for the thought-provoking discussions that have contributed to the development of the ideas presented in this paper. 

\bibliographystyle{unsrt}  
\bibliography{references}  






\end{document}